\documentstyle[11pt,moriond,epsfig]{article}

\begin{document}
\vspace*{4cm}
\title{OBSERVATIONS OF THE $\gamma$-RAY EMISSION ABOVE 250 GeV\\
FROM THE BLAZARS MARKARIAN~501 AND MARKARIAN~421\\
BY THE CAT CHERENKOV ATMOSPHERIC IMAGING TELESCOPE}

\author{F. PIRON, for the CAT collaboration}

\address{Laboratoire de Physique Nucl\'eaire des Hautes Energies\\
Ecole Polytechnique, route de Saclay, 91128 Palaiseau Cedex, France}

\maketitle
\abstracts{
The Very High Energy (VHE) $\gamma$-ray emission of the closest BL~Lacert{\ae} objects Markarian~501
and Markarian~421 has been observed by the CAT telescope in 1997 and 1998.
In 1997 Mrk~501 exhibited a remarkable series of flares, with a VHE emission peaking above
$250\:\mathrm{GeV}$. The source showed correlated emissions in the X-ray and VHE $\gamma$-ray bands, together with
intensity-spectral hardness correlation in the latter energy range. During small flares in 1998, Mrk~421 became the
second extragalactic source detected by CAT. Its spectral properties are compared to those of Mrk~501.
Theoretical implications for jet astrophysics are briefly discussed.}

\section{Introduction: the blazar family}
At the beginning of the 90's, the E{\small GRET} detector, operating onboard the {\it Compton Observatory}, definitively
opened the field of high energy astrophysics by revealing that most extragalactic strong $\gamma$-ray emitters were
blazars~\cite{VonMontigny}. The blazar family of Active Galactic Nuclei includes radio-loud objects such as
BL~Lacert{\ae} and Flat-Spectrum Radio Quasars~\cite{Urry95}. Besides a central engine, supposed to be a supermassive black hole
surrounded by an accretion disk, blazars show remarkable characteristics concerning the production
of high energy particles. Their emission is mainly non-thermal, with a featurless optical continuum, weak emission lines,
and strong optical polarization. Their radio jets,
made of an ultra-relativistic magnetized plasma with a bulk Lorentz factor of $\sim$10, are collimated and close to
the line of sight, giving rise to a strong Doppler boosting of the observed fluxes. In many cases, they exhibit
apparent superluminal motion, as revealed by radio interferometry observations at the parsec scale.
Their $\gamma$-radiation power often dominates the overall power emitted by the source,
and it must be produced in a small enough region to explain rapid variability time scales as short as a few hours or
less~\cite{Gaidos}. Currently, the base of the jets ($\sim$0.01pc from the black hole) is believed to
be the site of $\gamma$-ray production, whose detection by ground-based telescopes extends at least up to
$\sim20\:\mathrm{TeV}$~\cite{Aharonian1}. In the last few years, coordinated multiwavelength campaigns
have increased our understanding of blazar jet structure and of the mechanisms of energy extraction in the
surrounding of the black hole (see Ref.~7 for a recent review); although the origin of the jets is still uncertain, these
observations gave new insight into their particle content (hadronic or leptonic) and emission processes at the
sub-parsec scale.

\section{The CAT detector: characteristics and analysis method}
The C{\small AT} (Cherenkov Array at Th{\'e}mis) telescope~\cite{Barrau} started operation in Autumn 1996 on the site of
the former solar plant Th{\'e}mis (French Pyrenees). Through its 17.8m$^2$ mirror, it records Cherenkov
flashes due to VHE atmospheric showers. The fine grain of its camera (600 pixels with a 4.8$^\circ$ full field of
view), combined with fast electronics, allows a relatively low threshold ($250\:\mathrm{GeV}$ at Zenith) as well as an
accurate analysis~\cite{LeBohec} of the resulting images, which is necessary to distinguish between $\gamma$
and hadron-induced showers. Good discrimination is achieved by looking at the shape and the
orientation of the images: 
since $\gamma$-ray images are rather thin and ellipsoidal while hadronic images are more irregular, a first cut is applied
on the probability given by the fit of a semi-analytical model of mean electromagnetic showers in order to retain the
images with a ``$\gamma$-like'' shape. Then, since $\gamma$-ray images are expected to point towards the source
position in the focal plane whereas cosmic-ray directions are isotropic, images whose pointing angle~\footnote{For
point-like sources, the pointing angle is defined in the focal plane as the angle, at the image
barycentre, between the actual source position and the image's origin reconstructed by the
fit (resolution per event $\sim$0.1$^\circ$).} is too large are also rejected. As a result, this procedure rejects 99.5\% of
hadronic events while keeping 40\% of $\gamma$-ray events. A source like the Crab nebula, which is generally considered as
the standard candle for VHE $\gamma$-ray astronomy, can be detected at a 4.5$\sigma$ level in one hour and localised within
1' to 2'.
\section{CAT observations of Mrk~501 and Mrk~421}
\subsection{Markarian~501}\label{subsec501}
\vspace{-1.cm}
\begin{figure}[h]
\hbox{
\epsfig{file=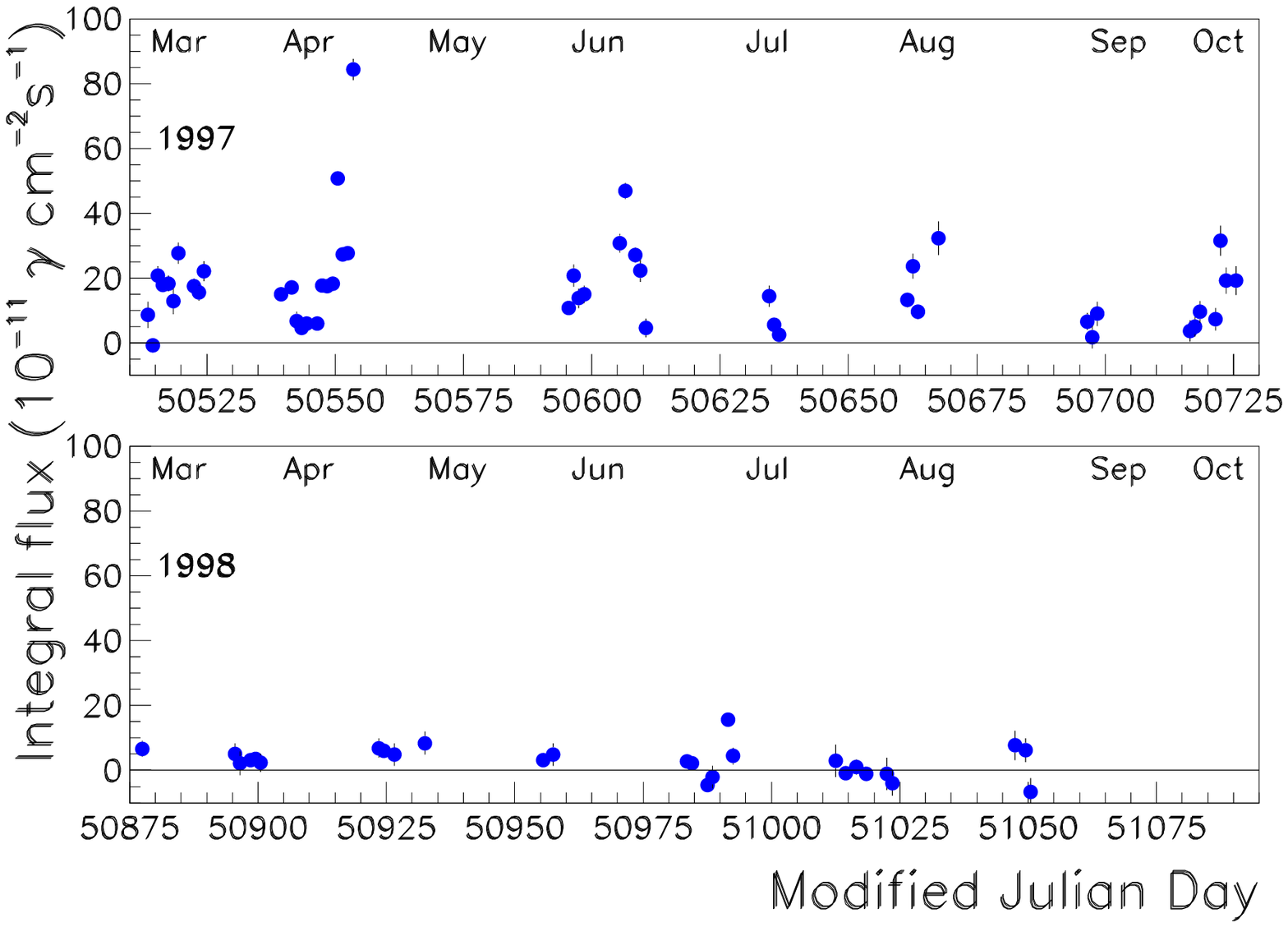,width=7cm,height=6cm}
\epsfig{file=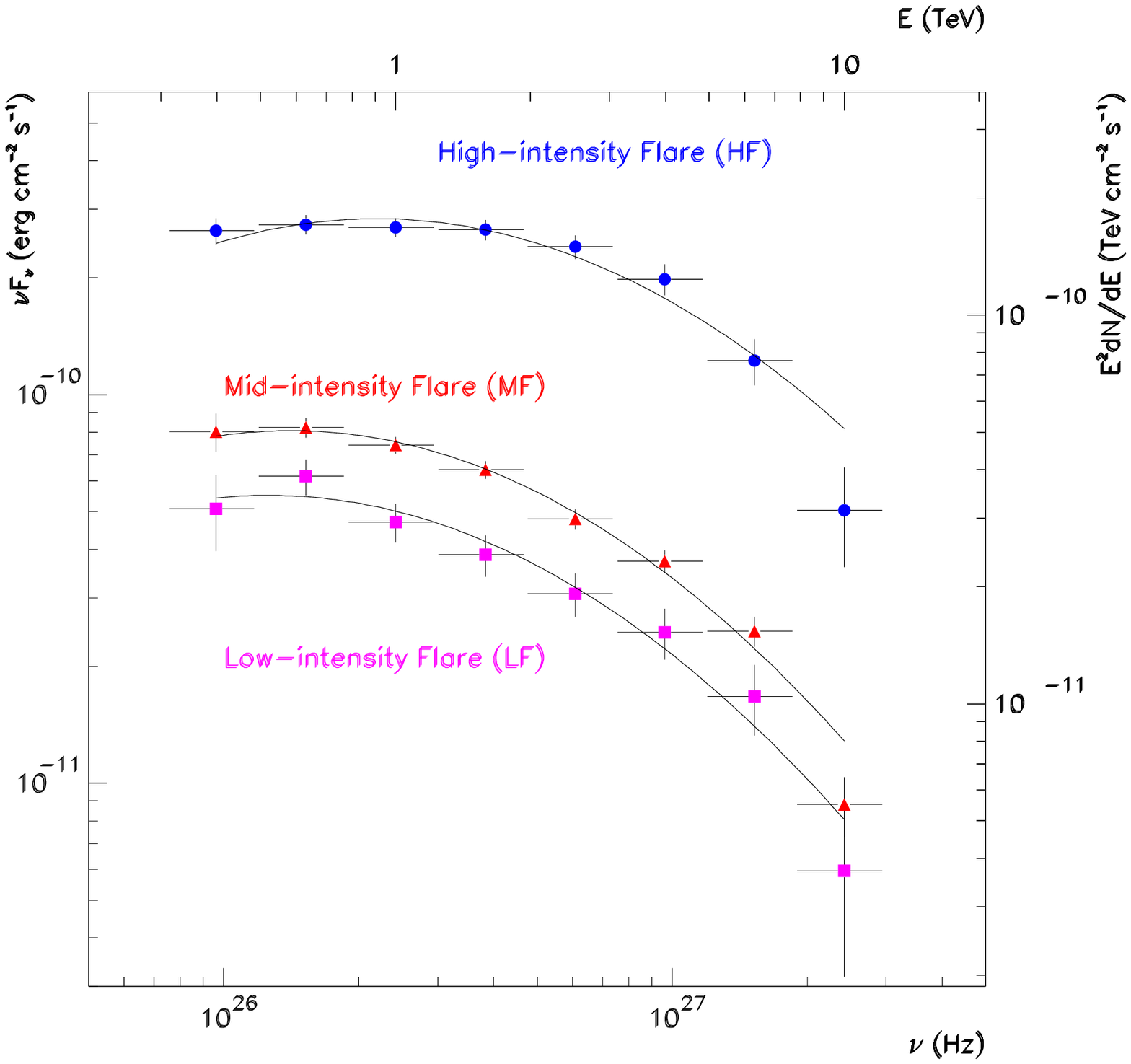,width=7cm,height=6cm}
}
\vspace{-.4cm}
\caption{
{\it (a)} Mrk~501 nightly integral flux above $250\:\mathrm{GeV}$ in 1997 and 1998;
{\it (b)} Mrk~501 VHE SED between $330\:\mathrm{GeV}$ and $13\:\mathrm{TeV}$ for high, mid and low flaring activity in 1997.
}
\label{501cl3nfn}
\end{figure}
\begin{figure}[tbh]
\hbox{
\epsfig{file=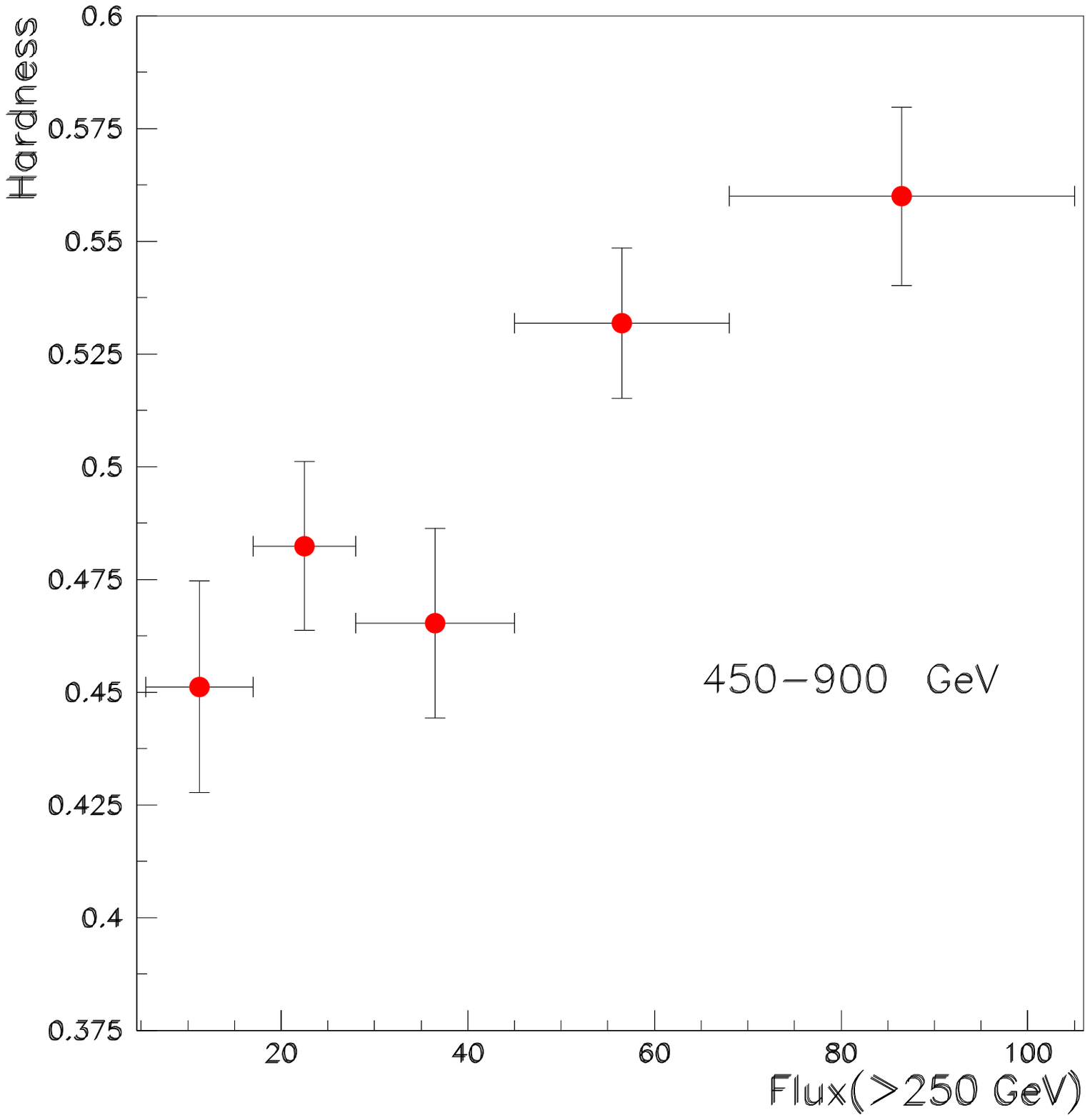,width=7cm,height=6cm}
\epsfig{file=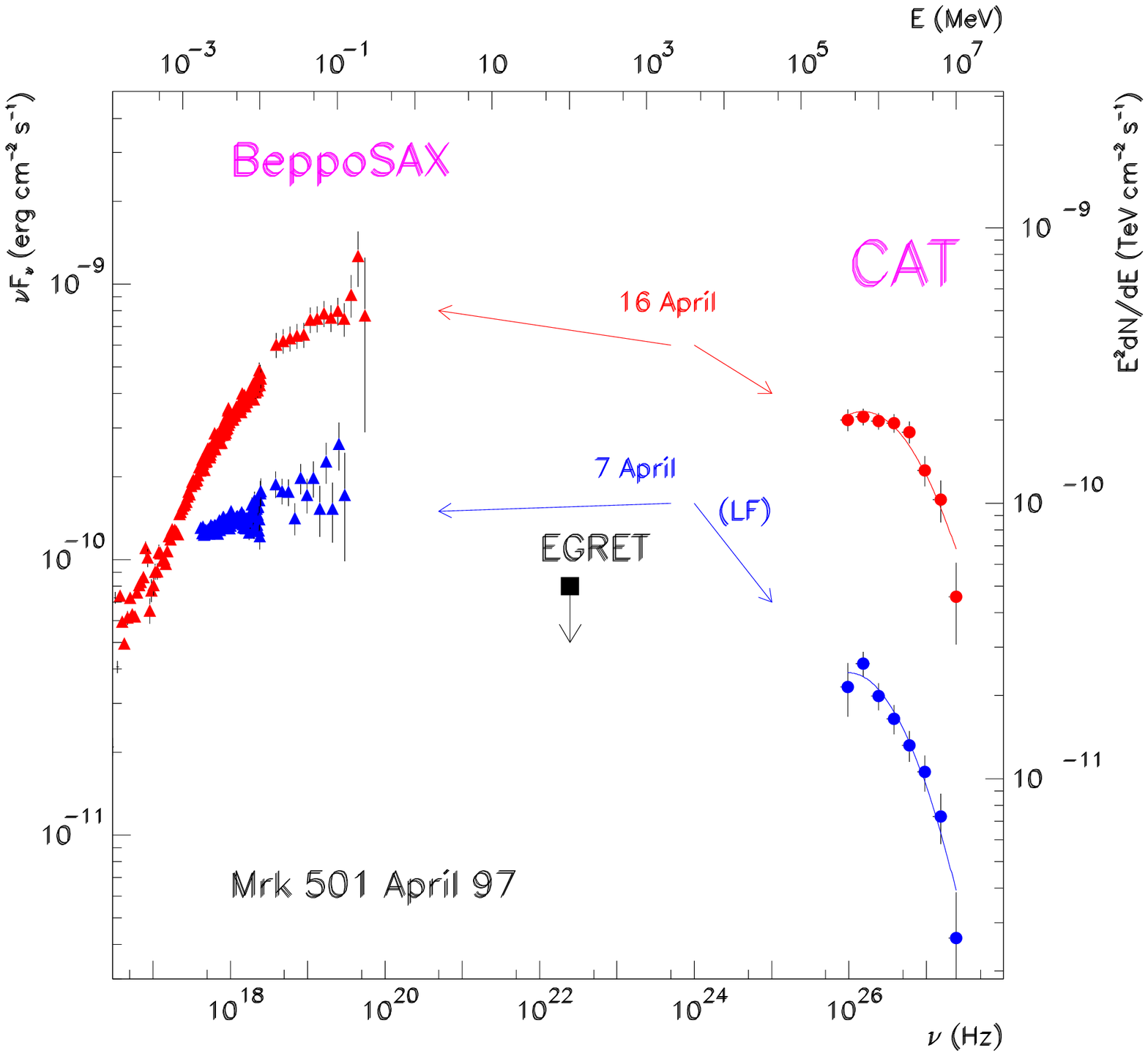,width=7cm,height=6cm}
}
\vspace{-.5cm}
\caption{
{\it (a)} Hardness-ratio ($HR=\frac{N_{E>900\:\mathrm{GeV}}}{N_{E>450\:\mathrm{GeV}}}$) {\it vs.} source intensity
(given as the integral flux above $250\:\mathrm{GeV}$ in units of $10^{-11}\:\mathrm{cm^{-2}s^{-1}}$);
{\it (b)} Mrk~501 X-ray and VHE spectra for April $7^{\mathrm{th}}$ and $16^{\mathrm{th}}$.
%The LF subset of CAT data was used to represent the spectrum of April $7^{\mathrm{th}}$.
The EGRET upper limit corresponds to observations between April $9^{\mathrm{th}}$ and $15^{\mathrm{th}}$.
}
\label{501hrbeppo}
\end{figure}

In 1997 Mrk~501 underwent a dramatic increase in intensity (Fig.~\ref{501cl3nfn}a), reaching $\sim$8 times
the level of the Crab nebula on April, 16$^{\mathrm{th}}$. The signal-to-noise ratio for this night
is 2.7, corresponding to a $\gamma$-ray beam with only 30\% contamination, and to the most powerful flare
ever recorded in VHE $\gamma$-ray astronomy~\cite{Protheroe}. The source remained very active during the whole year,
going down to a much lower mean flux in 1998.
The VHE spectral energy distribution (SED), derived for different flaring-activity states in 1997
(Fig.~\ref{501cl3nfn}b), shows a significant curvature which is now well confirmed by different ground-based
experiments~\cite{Aharonian1,Krennrich}. The peak $\gamma$-energy is found to lie just above the C{\small AT}
threshold, and it seems to shift towards higher energies as the flux increases: this tendancy can be seen in
Fig.~\ref{501cl3nfn}b. To check this spectral variability by a more robust method, the hardness ratio has been computed
for five different-level intensities:
the correlation observed in Fig.~\ref{501hrbeppo}a confirms the hardening of the VHE SED during flaring periods.
Fig.~\ref{501hrbeppo}b shows the broad-band SED of Mrk~501 for two flaring dates in April 1997: it
exhibits the two-bump structure typical of blazars, with a dip indicated by the contemporary E{\small GRET} upper-limit
point in the GeV energy range~\cite{Samuelson}. The obvious correlation of the X-ray emission from BeppoS{\small AX}
data~\cite{Pian} with TeV emission strongly suggests that the same particle
population is responsible for emission in both energy-ranges, i.e. it supports the picture given by leptonic
models~\cite{Ghisellini}, in which an energetic electron beam propagating in the magnetized plasma jet produces
X-rays through synchrotron radiation as well as VHE $\gamma$-rays through inverse Compton scattering of low-energy
photons.\\
\vspace{-0.8cm}
\subsection{Markarian~421}
\vspace{-0.8cm}
\begin{figure}[tbh]
\hbox{
\epsfig{file=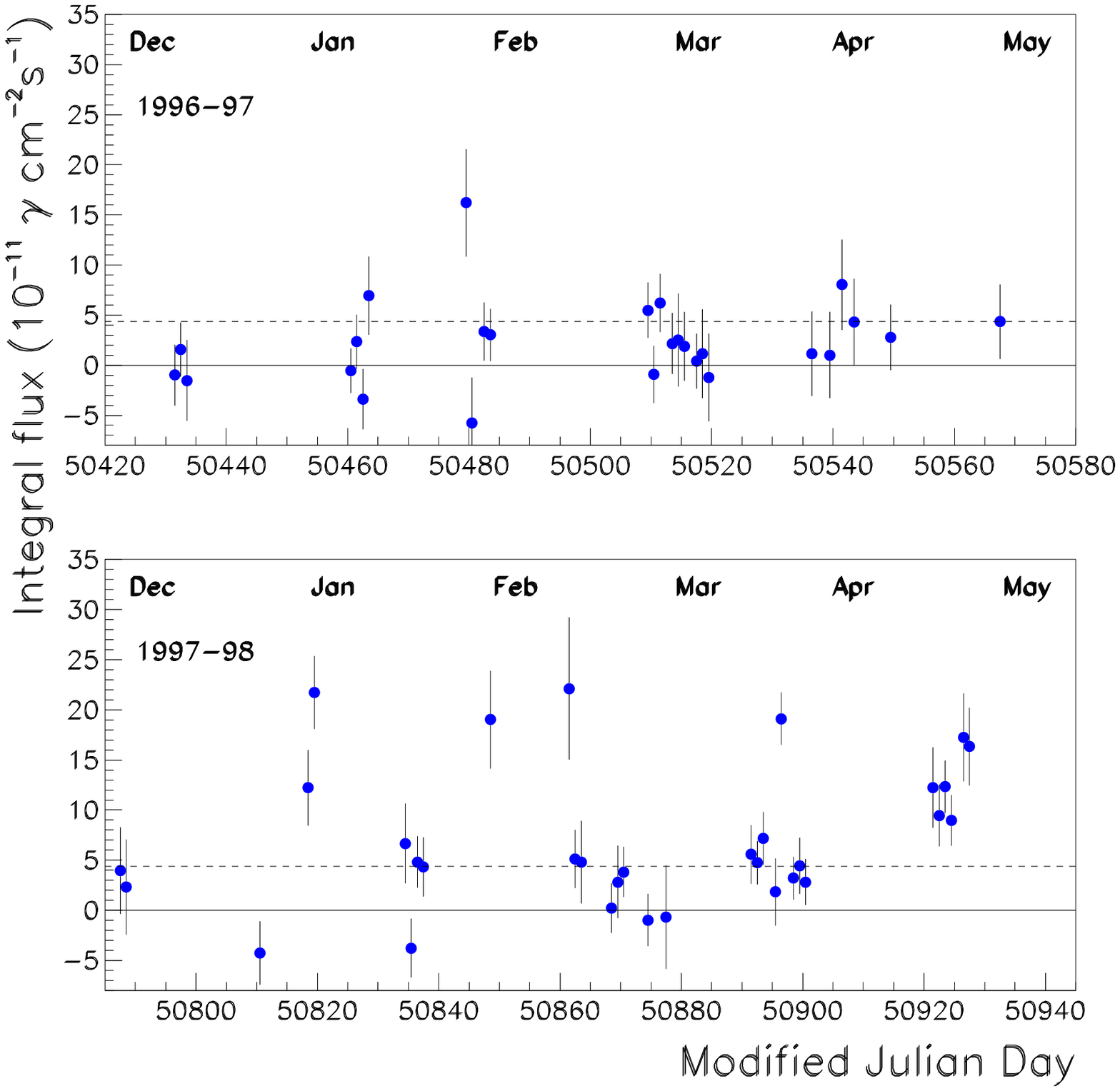,width=7cm,height=6cm}
\epsfig{file=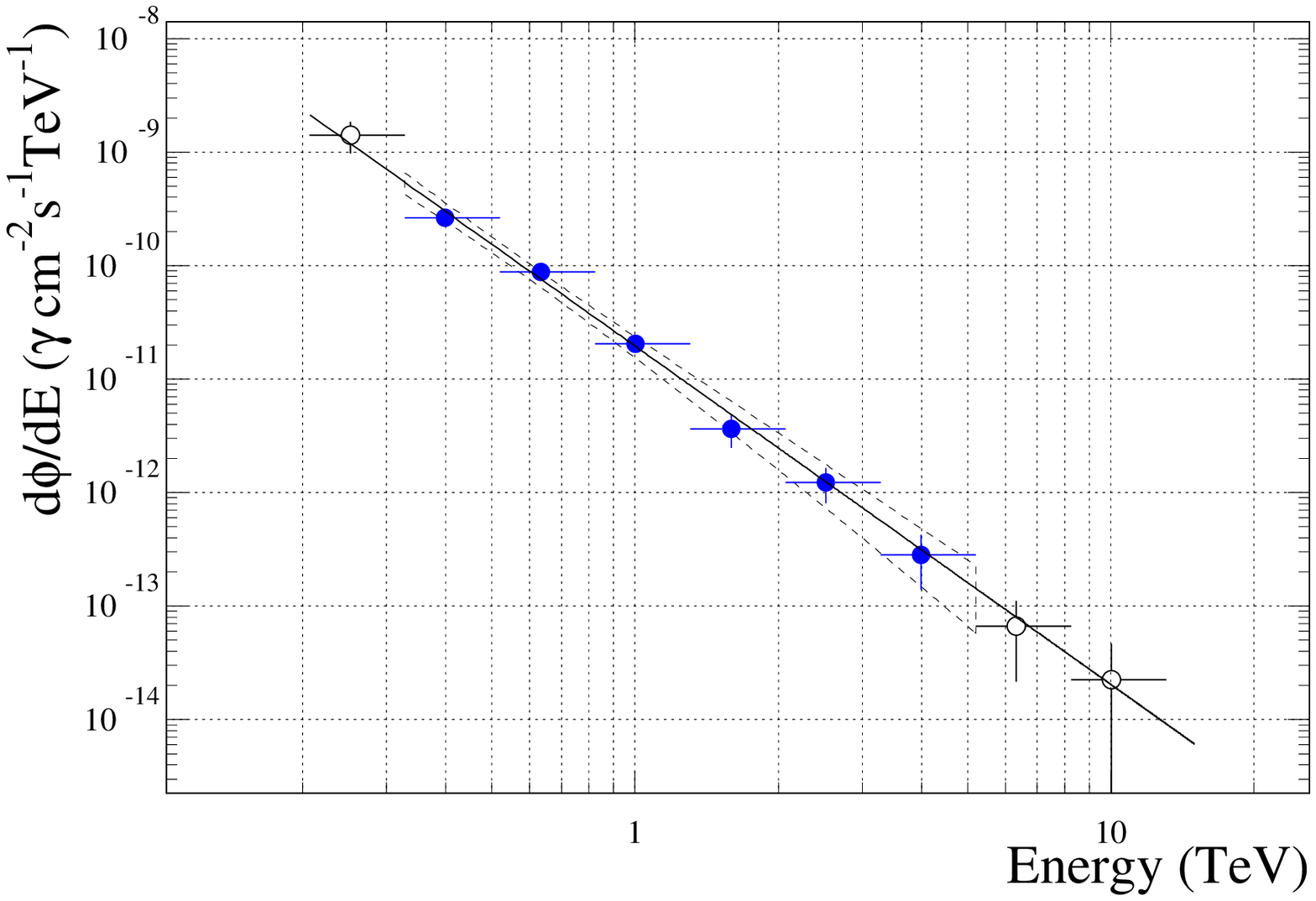,width=7cm,height=6cm}
}
\vspace{-.1cm}
\caption{
{\it (a)} Mrk~421 nightly integral flux above $250\:\mathrm{GeV}$
between December 1996 and May 1998. The dashed line represents the mean flux over the two years;
{\it (b)} Differential flux of Mrk~421 for the flaring periods in 1998.
The full line represents the power-law parameters yielded by a likelihood method,
in which only the energy bins shown with fullfilled circles were used.
The shape of the $2\sigma$ error ``wedge'' reflects the correlation between the parameters.
%This method gives the $2\sigma$ error ``box'', whose shape reflects the correlation
%between the parameters.
%This latter vanishes at the decorrelation energy
%$\mathrm{E_d}=580\:\mathrm{GeV}$.
}
\label{421clsp}
\end{figure}
The mean $\gamma$-rate of Mrk~421 between December 1996 and May 1998 is found to be of the order of a third of
that of the Crab nebula, with a total significance reaching the 7.6$\sigma$ level. This makes it
the second extragalactic source, after Mrk~501, detected by C{\small AT}~\cite{Piron}. The emission of Mrk~421 changed significantly between
1996-97 and 1997-98 (Fig.~\ref{421clsp}a): almost quiet in the first period,
%($\langle{\Phi}_{>250\:\mathrm{GeV}}\rangle = 1.94 \pm 0.65 \times 10^{-11}\:\mathrm{cm^{-2}s^{-1}}$),
the source showed small bursts in the second period together with a higher mean activity.
%($\langle{\Phi}_{>250\:\mathrm{GeV}}\rangle = 6.05 \pm 0.54 \times 10^{-11}\:\mathrm{cm^{-2}s^{-1}}$).
The energy spectrum derived from $330\:\mathrm{GeV}$ to $5.2\:\mathrm{TeV}$ for the 1998 flaring periods is well
represented by a simple power law (Fig.~\ref{421clsp}b) with a differential spectral index of $2.96 \pm 0.13$,
which confirms the absence of any obvious spectral curvature~\cite{Aharonian2,Krennrich}.
In the framework of leptonic models~\cite{Ghisellini}, which succesfully explain the
Mrk~501 broad-band SED in 1997~\cite{Djannati} (see \ref{subsec501}),
this result implies that the peak energy of the inverse-Compton contribution to Mrk~421's spectrum
is significantly lower than the C{\small AT} detection threshold.
This is not surprising since the corresponding synchrotron peak is lower than that of Mrk~501 in 1997, and
since we have seen that leptonic models predict a strong correlation between X-rays and $\gamma$-rays. In
fact, such a correlation was directly observed on Mrk~421 in Spring 1998, during a coordinated
observation campaign involving ground-based Cherenkov imaging telescopes (Whipple, H{\small EGRA},
and C{\small AT}) and the A{\small SCA} X-ray satellite\cite{Takahashi99}.
\section{Conclusion}
Since Mrk~501 and Mrk~421 lie at the the same redshift ($\sim$0.03), spectral differences between them must
be intrinsic and not due, in particular, to absorption by the diffuse infrared background radiation. This allows
direct and relevant comparison of their spectral properties. The correlation between X-ray and $\gamma$-ray
emissions is now proven for both sources, supporting the simple and most natural scenario given by leptonic
models~\cite{Ghisellini} for the origin of blazar TeV flares, in which a single leptonic population is injected into
the radio jets and produces correlated X-ray synchrotron and VHE $\gamma$-ray inverse Compton radiations.
C{\small AT} observations complete this picture with some evidence of spectral variability of Mrk~501 in 1997,
suggesting that this leptonic population is responsible for the hardening of the entire high-energy part of the
electromagnetic spectrum during flares. To date, the Mrk~421 peak $\gamma$-energy has always remained well below the
C{\small AT} threshold, precluding a more accurate spectral study. Therefore, testing VHE spectral variability as a general
feature of blazars requires more multiwavelength observations with a large dynamic range in intensity.


\section*{References}
\begin{thebibliography}{99}
\bibitem{Aharonian1}Aharonian, F.A., {\it et al}, {\it A\&A} {\textbf{349}}, 11 (1999).%HEGRA average 501 1997
\bibitem{Aharonian2}Aharonian, F.A., {\it et al}, {\it A\&A} {\textbf{350}}, 757 (1999).%HEGRA 421 1997-98
\bibitem{Barrau}Barrau, A., {\it et al}, {\it Nucl. Instr. Meth.} A {\textbf{416}}, 278 (1998).
\bibitem{Djannati}Djannati-Ata\"{\i}, A., {\it et al}, {\it A\&A} {\textbf{350}}, 17 (1999).
\bibitem{Gaidos}Gaidos, J.A., {\it et al}, {\it Nature} {\textbf{383}}, 319 (1996).
\bibitem{Ghisellini}Ghisellini, G., Maraschi, L., and Dondi, L., {\it A\&AS} {\textbf{120}}, 503 (1996).
\bibitem{Hoffman99}Hoffman, C.M., {\it et al}, {\it Rev. Mod. Phys.} {\textbf{71}}, vol.4, p.897 (1999).
\bibitem{Krennrich}Krennrich, F., {\it et al}, {\it ApJ} {\textbf{511}}, 149 (1999).
\bibitem{LeBohec}Le Bohec, S., {\it et al}, {\it Nucl. Instr. Meth.} A {\textbf{416}}, 425 (1998).
\bibitem{Pian}Pian, E., {\it et al}, {\it ApJL} {\textbf{492}}, 17 (1998).
\bibitem{Piron}Piron, F., {\it et al}, {\it Proc. XXVI ICRC}  {\textbf 3}, 326 (Salt-Lake City, 1999).
\bibitem{Protheroe}Protheroe, R. J., {\it et al}, {\it Proc. XXV ICRC}  {\textbf 8}, 317 (Durban, 1997).
\bibitem{Samuelson}Samuelson, F., {\it et al}, {\it ApJL} {\textbf{501}}, 17 (1998).
\bibitem{Takahashi99}Takahashi, T., {\it et al}, {\it APh} {\textbf{11}}, 177 (1999).
\bibitem{Urry95}Urry, C.M., and Padovani, P., {\it Publ. Astr. Soc. Pacif.} {\textbf{107}}, 803 (1995).
\bibitem{VonMontigny}Von Montigny, C., {\it et al}, {\it ApJ} {\textbf{440}}, 525 (1995).
\end{thebibliography}
\end{document}